# Extended Harmonic Map Equations and the Chaotic Soliton Solutions

Gang Ren[1,*] and Yi-Shi Duan[2,*]

[1] The Molecular Foundry, Lawrence Berkeley National Laboratory, Berkeley CA 94720, USA
[2] The Department of Physics, Lanzhou University, Lanzhou, Gansu 730000, China

* To whom correspondence should be addressed: G. Ren, Lawrence Berkeley National Laboratory, the Molecular Foundry, 1 Cyclotron Road, Berkeley, CA 94720, USA, E-mail: gren@lbl.gov or Y.S. Duan, The Department of Physics, Lanzhou University, Lanzhou, Gansu 730000, China, E-mail: ysduan@lzu.edu.cn

**Abstract**

In this paper, the theory of harmonic maps is extended. The soliton or travelling wave solutions of Euler's equations of the extended harmonic maps are studied. In certain cases, the chaotic behaviours of these partial equations can be found for the particular case of the metrics and the potential functions of the extended harmonic equations.

**Introduction**

All physical processes are described by differential equations. Over the last several decades, many nonlinear ordinary differential equations have been found to exhibit chaotic behaviours, such as the nonlinear equation for a parametrically excited pendulum given by[1]

$$\frac{d^2x}{dt^2} + k\frac{dx}{dt} + [\alpha + \beta Cos(\varpi t)]Sin(x) = 0 \tag{1}$$

where k, α and β are control parameters, and the equation for a harmonic system in periodic fields [2] given by

$$\frac{d^2x}{dt^2} + k\frac{dx}{dt} - \beta x + \alpha x^3 = bCos(\varpi t) \tag{2}$$

where k, α and β are control parameters. There is also the Lorentz model[3],

$$\frac{dx}{dt} = -\alpha(x-y); \frac{dy}{dt} = -xy - \beta x - y; \frac{dz}{dt} = xy - kx \tag{3}$$

where k, α and β are control parameters.

Over the last several decades, many nonlinear ordinary differential equations possessing chaotic behaviours have been discovered[4]. To date, the study of chaotic behaviours using the space distribution of nonlinear partial differential equations has been rare and is currently only in its initial stage. In this paper, we use the extended theory of harmonic maps to show that chaotic behaviours exist for a certain class of nonlinear partial differential equations.

Three decades ago, Duan proposed a method to solve the Euler's equations of harmonic maps (HM)[5], in which the solution of the Euler's equations of HM was found by solving the

Laplace-Beltrami's equation and the geodesic equations. Using this method, the general solutions of Einstein equations in general relativity have been discussed. From the viewpoint of physics, the traditional theory of harmonic maps contains only the kinetic energy term. For a wider application of this method, the Lagrangian of the traditional HM theory must be supplemented with a potential energy term[6]. This extended HM theory is helpful for studying the travelling wave or soliton solutions of some types of nonlinear partial differential equations.

Here, we first establish an extended theory of harmonic maps and discuss the travelling wave or soliton solution of the Euler's equations of the extended HM. We then use the extended theory of HM and study two special cases for which the chaotic behaviours of some types of partial differential equations are found. The theory and method could be helpful for studying the existence of soliton solutions and the chaotic behaviours of more types of partial equations, possibly providing a further link between the extended HM theory and quantum entanglement via common classical chaos [7].

**The Extended Theory of Harmonic Maps**
The theory of harmonic maps [8-10] became an important branch of mathematical physics many years ago and has been applied to a wide variety of problems in mathematics and theoretical physics. In this section, we first describe the formulation of the extended theory of harmonic maps.

Let M and N be two Riemannian manifolds with local coordinates $x^\mu$ ($\mu =1, 2, \ldots, m$) on M and local coordinates $\Phi^A$ (A=1, 2, …, n) on N. The metrics on M and N are denoted by

$$dl^2 = g^{\mu\nu}(x)dx^\mu dx^\nu \qquad \dim(M)=m$$
$$d\tilde{L}^2 = G_{AB}(\Phi)d\Phi^A d\Phi^B \qquad \dim(N)=n$$

respectively. A mapping

$$\Phi: M \to N$$
$$x \to \Phi$$

is called an extended harmonic map if it satisfies the Euler's equation resulting from the variational principle $\delta I = 0$, using the action

$$I = \int d^n x \sqrt{g} \left[ -\frac{1}{2} g^{\mu\nu} \partial_\mu \Phi^A \partial_\nu \Phi^B G_{AB}(\Phi) + V(\Phi) \right] \qquad (4)$$

where $\partial_\mu = \frac{\partial}{\partial x^\mu}$, V($\Phi$) is the potential function of $\Phi^A$, and

$$V(\Phi) = V(\Phi^1, \Phi^2, \ldots, \Phi^n) \,.$$

It is obvious that the traditional HM corresponds to the case of $V(\Phi) = 0$. The conditions for a map to be harmonic are given by the Euler's equations

$$\frac{\partial L}{\partial \Phi^A} - \partial_\mu \frac{\partial L}{\partial \partial_\mu \Phi^A} = 0 \qquad A=1, 2, \ldots, n \qquad (5)$$

where
$$L = -\frac{1}{2}g^{\mu\nu}\partial_\mu\Phi^A\partial_\nu\Phi^B G_{AB}(\Phi)\sqrt{g} + V(\Phi)\sqrt{g}. \tag{6}$$

By substituting (6) into (5), we can obtain the Euler's equations of extended HM given by

$$\frac{1}{\sqrt{g}}\partial_\mu(\sqrt{g}g^{\mu\nu}\partial_\nu\Phi^A) + g^{\mu\nu}\Gamma^A_{BC}\partial_\mu\Phi^B\partial_\nu\Phi^C + G^{AB}\frac{\partial V(\Phi)}{\partial\Phi^B} = 0 \tag{7}$$

where $\Gamma^A_{BC}$ are the Christoffels symbols on manifold N given by

$$\Gamma^A_{BC} = \frac{1}{2}G^{AD}[\frac{\partial G_{BD}}{\partial\Phi^C} + \frac{\partial G_{CD}}{\partial\Phi^B} - \frac{\partial G_{BC}}{\partial\Phi^D}]. \tag{8}$$

To study a special type of solution of partial differential equation (7), following reference [5], we investigate the case where $\Phi^A$ (A=1, 2, …, n) are functions solely of the argument σ, and σ is a function of $x^\mu$ on the manifold M:

$$\Phi^A = \Phi^A(\sigma) \tag{9}$$
$$\sigma = \sigma(x); \tag{10}$$

By substituting (9) into (7), equation (7) can be written as

$$\frac{1}{\sqrt{g}}\partial_\mu(\sqrt{g}g^{\mu\nu}\partial_\nu\sigma)\frac{d\Phi^A}{d\sigma} + G^{AB}\frac{\partial V(\Phi)}{\partial\Phi^B} + (\frac{d^2\Phi^A}{d\sigma^2} + \Gamma^A_{BC}\frac{d\Phi^B}{d\sigma}\frac{d\Phi^C}{d\sigma})g^{\mu\nu}\partial_\mu\sigma\partial_\nu\sigma = 0 \tag{11}$$

It is easy to see that if the function $\sigma = \sigma(x)$ satisfies the Laplace-Beltrami equations

$$\frac{1}{\sqrt{g}}\partial_\mu(\sqrt{g}g^{\mu\nu}\partial_\nu\sigma) = 0 \tag{12}$$

equation (11) will take the form

$$\frac{d^2\Phi^A}{d\sigma^2} + \Gamma^A_{BC}\frac{d\Phi^B}{d\sigma}\frac{d\Phi^C}{d\sigma} = -\frac{1}{f}G^{AB}\frac{\partial V(\Phi)}{\partial\Phi^B} \tag{13}$$

where
$$f = g^{\mu\nu}\partial_\mu\sigma\partial_\nu\sigma \tag{14}$$

is a scalar function on the manifold M. Because $f$ is not the function of $V(\Phi) = fU(\Phi)$, equation (13) can also be written as

$$\frac{d^2\Phi^A}{d\sigma^2} + \Gamma^A_{BC}\frac{d\Phi^B}{d\sigma}\frac{d\Phi^C}{d\sigma} = -G^{AB}\frac{\partial U(\Phi)}{\partial \Phi^B} \qquad (15)$$

Equation (15) can viewed as the geodesic equation of a particle on the Riemannian manifold N subjected to an external force

$$F^A = -G^{AB}\frac{\partial U(\Phi)}{\partial \Phi^B} \qquad (16)$$

This will become equivalent to the usual geodesic equations on the manifold N if the external force $F^A = 0$ (A=1, 2, ..., n).

A more general form of equation (15) can be written because the existence and uniqueness of the solution of the above equation has been discussed [11]. Therefore, generally, a solution of (15) exists when the initial conditions are given by $\Phi^A|_{\sigma=\sigma_0}$ and $\frac{d\Phi^A}{d\sigma}|_{\sigma=\sigma_0}$.

From (11), (12) and (14), we see that if $\Phi^A$ satisfies the geodesic equations with an external force on the manifold N, *i.e.,* (13), and $\sigma = \sigma(x)$ satisfies the Laplace-Beltrami equations on the manifold M (12), the functions $\Phi^A = \Phi^A(\sigma(x))$ (A=1, 2,..., n) will be the special solutions of the Euler's equations of the extended HM (7). In the following, we use $\sqrt{-g}$ instead of $\sqrt{g}$ if the manifold M is the pseudo-Riemannian space-time. Now, we show that there exist travelling wave or soliton solutions for partial differential equations (7). It can be proven that a solution of the Laplace-Beltrami equation (12) will be

$$\sigma = k_\mu x^\mu \qquad (17)$$

where $k_\mu$ ($\mu$ =0, 1,..., m-1) are components of a constant vector and $x^0 = t$ is time. Substituting (17) into (12), we find

$$\frac{1}{\sqrt{-g}}\partial_\mu(\sqrt{-g}g^{\mu\nu})k_\nu = 0 \qquad (18)$$

and choosing Fock's harmonic coordinates that always exist in pseudo-Riemannian M [12], *i.e.,*

$$\frac{1}{\sqrt{-g}}\partial_\mu(\sqrt{-g}g^{\mu\nu}) = 0 \quad, \qquad (19)$$

equation (18) is evidently satisfied. This means that $\sigma = k_\mu x^\mu$ is indeed a solution of equation (12) and that the scalar function $f$ defined by (13) is of the form

$$f = g^{\mu\nu}k_\mu k_\nu \quad. \qquad (20)$$

Therefore, the solution of the equations of motion (15) on the manifold N with given initial conditions can be formally expressed as

$$\Phi^A = \Phi^A(\sigma) = \Phi^A(k_\mu x^\mu) = \Phi^A(\vec{k}\cdot\vec{x} - k_0 t) \quad (A=1, 2,\ldots,n) \tag{21}$$

The above expression shows that the partial differential equations of the extended HM (7) in pseudo-Riemannian space-time possess solutions.

In the case that the manifold M is the pseudo-Euclidean space-time

$$g^{\mu\nu} = \eta^{\mu\nu}$$

where $\eta^{\mu\nu}$ is a Lorentz metric, equation (12) is simplified to

$$\Box\sigma = \eta^{\mu\nu}\partial_\mu\partial_\nu\sigma = 0 \tag{22}$$

which has an obvious solution $\sigma = k_\mu x^\mu$, and the function $f$ defined by (20) is a constant

$$f = \eta^{\mu\nu} k_\mu k_\nu \tag{23}$$

In this case, the extended HM equation (7) becomes

$$\Box\Phi^A + \Gamma^A_{BC}\partial_\mu\Phi^B\partial_\nu\Phi^C \eta^{\mu\nu} + G^{AB}\frac{\partial V(\Phi)}{\partial \Phi^B} = 0 \tag{24}$$

**Chaotic soliton solutions of the extended HM equations**

In the previous section, it has been noted that there exist travelling wave or soliton solutions of the partial differential equations of the extended HM (7) in the pseudo-Riemannian space-time. The form (shape) of travelling wave or soliton $\Phi^A = \Phi^A(\sigma)$ is determined by the solution of nonlinear ordinary differential equations (14).

In the following, we study the simple case that the M is the pseudo-Euclidean space-time and N is a 2-dimensional manifold with coordinates $\{\Phi^1, \Phi^2\}$. As in the above section, we suppose that $\Phi^1$ and $\Phi^2$ are functions of an argument $\sigma = \sigma(x) = k_\mu x^\mu$ and further assume that $\Phi^2 = \sigma$, i.e.,

$$\Phi^1 = \Phi(\sigma); \quad \Phi^2 = \sigma; \quad \sigma = k_\mu x^\mu \tag{25}$$

Then, equation (14) can be expressed as

$$\frac{d^2\Phi}{d\sigma^2} + \Gamma^1_{11}\left(\frac{d\Phi}{d\sigma}\right)^2 + 2\Gamma^1_{12}\left(\frac{d\Phi}{d\sigma}\right) + \Gamma^1_{22} + \frac{1}{f}\left(G^{11}\frac{\partial V(\Phi,\sigma)}{\partial \Phi} + G^{12}\frac{\partial V(\Phi,\sigma)}{\partial \sigma}\right) = 0 \tag{26}$$

$$\Gamma_{11}^2(\frac{d\Phi}{d\sigma})^2 + 2\Gamma_{12}^2(\frac{d\Phi}{d\sigma}) + \Gamma_{22}^2 + \frac{1}{f}(G^{21}\frac{\partial V(\Phi,\sigma)}{\partial \Phi} + G^{22}\frac{\partial V(\Phi,\sigma)}{\partial \sigma}) = 0.$$

(27)

Here, we have denoted the potential function as

$$V(\Phi^1,\Phi^2) = V(\Phi,\sigma)$$

Eliminating the term $(\frac{d\Phi}{d\sigma})^2$ from the differential equations (26) and (27), we obtain

$$\frac{d^2\Phi}{d\sigma^2} + 2(\Gamma_{12}^1 - \frac{\Gamma_{11}^1\Gamma_{12}^2}{\Gamma_{11}^2})\frac{d\Phi}{d\sigma} + (\Gamma_{22}^1 - \frac{\Gamma_{11}^1\Gamma_{21}^2}{\Gamma_{11}^2})$$
$$+ \frac{1}{f}[G^{11}\frac{\partial V(\Phi,\sigma)}{\partial \Phi} + G^{12}\frac{\partial V(\Phi,\sigma)}{\partial \sigma} - \frac{\Gamma_{11}^1}{\Gamma_{11}^2}(G^{21}\frac{\partial V(\Phi,\sigma)}{\partial \Phi} + G^{22}\frac{\partial V(\Phi,\sigma)}{\partial \sigma}] = 0$$

(28)

To make the equation have the same form as equation (2), we assume that the metrics on manifold N are diagonal and take the following form:

$$G_{11} = e^{\Phi+\sigma}; G_{22} = (k-1)e^{\Phi+\sigma}; G_{12} = G_{21} = 0$$

(29)

where k is a constant, and the potential function V(Φ, σ) takes the form

$$V(\Phi,\sigma) = fe^{\Phi+\sigma}u(\Phi,\sigma)$$

(30)

$$u(\Phi,\sigma) = \frac{\alpha}{2}\Phi^3 - \frac{3}{4}\alpha\Phi^2 - \frac{1}{2}[\beta - \frac{3}{2}\alpha]\Phi + \frac{1}{4}(\beta - \frac{3}{2}\alpha)\Phi$$
$$- \frac{2b}{\varpi^2+4}Cos(\varpi\sigma) - \frac{\varpi b}{\varpi^2+4}Sin(\varpi\sigma)$$

(31)

Substituting (8), (29), (30) and (31) into equation (28), we obtain

$$\frac{d^2\Phi}{d\sigma^2} + k\frac{d\Phi}{d\sigma} - \beta\Phi + \alpha\Phi^3 - bCos(\varpi\sigma) = 0; \sigma = k_\mu x^\mu$$

(32)

Comparing the above equation for $\Phi$ with equation (2), we find that they are of the same form. It has been shown that equation (2), *i.e.,* the equation of a harmonic system in a periodic field, possesses chaotic states that are characterized by the existence of a strange attractor in the phase space. This implies that equation (32) for $\Phi(\sigma)$ should possess the same chaotic states as (2). Because equation (32) is deduced from the special case of equation (14), it determines the form (shape) of soliton $\Phi(\sigma) = \Phi(k_\mu x^\mu)$, as discussed in the previous section. Therefore, the chaotic states of equation (32) mean that the form (shape) of solution determined by this equation exhibits chaotic behaviours, and we call this type of solution of (32) the chaotic soliton.

Substituting (8), (29), (30), (31) and $\Phi^1 = \Phi(\sigma)$; $\Phi^2 = \sigma = k_\mu x^\mu$ into (24), we find two partial differential equation for $\Phi$. Eliminating the quadric term $\eta^{\mu\nu}\partial_\mu\Phi\partial_\nu\Phi$ from these two equations, we have the partial differential equation

$$\Box\Phi + k\eta^{\mu\nu}\partial_\mu\Phi k_\nu + f[\alpha\Phi^3 - \beta\Phi - bCos(\varpi k_\mu x^\mu)] = 0 \qquad (33)$$

If we let $\Phi = \Phi(\sigma)$ and $\sigma = k_\mu x^\mu$ and substitute these equations into (33), it is easy to verify that equation (32) can be directly derived from (33). This means that equation (33) has the soliton solution $\Phi = \Phi(k_\mu x^\mu)$ and that equation (32) determines the form of the soliton $\Phi = \Phi(\sigma)$. From the above discussion, it is obvious that the partial differential equation (33) possesses a special type of a chaotic soliton solution.

In the following, we study another example, for which the function $u(\Phi,\sigma)$ in the potential function (29) is chosen as

$$u(\Phi,\sigma) = \frac{2}{5}\alpha Sin(\Phi) - \frac{1}{5}\alpha Cos(\Phi) - \frac{\beta}{2}[\frac{1+\varpi}{4+(1+\varpi)^2} - \frac{1-\varpi}{4+(1-\varpi)^2}]Sin(\varpi\sigma)Sin(\Phi)]$$
$$+ \beta[\frac{1}{4+(1+\varpi)^2} - \frac{1}{4+(1-\varpi)^2}]Sin(\varpi\sigma)Cos(\Phi)$$
$$+ \beta[\frac{1}{4+(1+\varpi)^2} + \frac{1}{4+(1-\varpi)^2}]Cos(\varpi\sigma)Sin(\Phi) \qquad (34)$$
$$- \frac{\beta}{2}[\frac{1+\varpi}{4+(1+\varpi)^2} + \frac{1-\varpi}{4+(1-\varpi)^2}]Cos(\varpi\sigma)Cos(\Phi)$$

Substituting (34) into (28), we obtain the following equation corresponding to equation (32):

$$\frac{d^2\Phi}{d\sigma^2} + k\frac{d\Phi}{d\sigma} + [\alpha + \beta Cos(\varpi\sigma)]\sin(\Phi) = 0 \qquad (35)$$

This is merely the nonlinear equation of a parametrically excited damped pendulum (1). Reference [4] has explored in detail the chaotic behaviours of this equation. The partial differential equation corresponding to (35) can be obtained using (24), (29), (30) and (34). That is,

$$\Box\Phi + k\eta^{\mu\nu}\partial_\mu\Phi k_\nu + f[\alpha + \beta Cos(\varpi k_\mu x^\mu)]Sin(\Phi) = 0 \qquad (36)$$

As discussed in the first example, we know that this partial differential equation also has chaotic soliton solutions. The chaotic behaviour of such solitons of equation (36) is determined by the time-evolution curve of the amplitude of the parametric pendulum model.

The two examples discussed above are both nonlinear damped oscillator systems that are the special cases of the Duffing equation [4].

$$\frac{d^2\Phi}{d\sigma^2} + k\frac{d\Phi}{d\sigma} + [p(\Phi) - g(\Phi,\sigma)] = 0 \quad . \tag{37}$$

Here, $g(\Phi,\sigma)$ is a function of $\Phi$ and $\sigma$, which is a periodic function of $\sigma$. $p(\Phi)$ is a nonlinear function of $\Phi$. In this general case, the function $u(\Phi,\sigma)$ in (31) is the solution of the following equation

$$2u(\Phi,\sigma) + \frac{\partial u(\Phi,\sigma)}{\partial \Phi} + \frac{\partial u(\Phi,\sigma)}{\partial \sigma} = p(\Phi) - g(\Phi,\sigma) \tag{38}$$

Using (38), we can derive the following partial differential equation with chaotic soliton solutions

$$\Box\Phi + k\eta^{\mu\nu}\partial_\mu\Phi k_\nu + [p(\Phi) - g(\Phi,\sigma)] = 0 \tag{39}$$

with the form (shape) of the solitons determined by the time-evolution curve of the amplitude described by the diffusion equation (37)

**Conclusion**

The theory presented in this paper is helpful for studying the existence of the soliton and chaotic soliton solutions for certain types of nonlinear partial differential equations. Using the Lorentz model[3], which is the standard model for the study of chaos theory, we have deduced a system of nonlinear partial differential equations. These equations possess chaotic travelling wave or soliton solutions with the space distribution.

**Acknowledgments:** We thank Mr. Wencheng Ren to preserve the original results and manuscript. Work at the Molecular Foundry was supported by the Office of Science, the Office of Basic Energy Sciences, and the U.S. Department of Energy, under Contract No. DE-AC02-05CH11231.

**Author Contributions:** This project was initiated by GR and YD., GR calculated the solution, and YS validated the solution. GR drafted the initial manuscript, which was revised by YD.

**Additional Information**

**Competing financial interests**
The author(s) declare no competing financial interests.